\documentclass[twocolumn,amsmath,amssymb,
aps,showpacs,graphicx]{revtex4-2}

\usepackage{graphicx}
\usepackage{dcolumn}
\usepackage{bm}
\usepackage{lineno, hyperref}
\usepackage{natbib}
\begin{document}

\title{Excitonic correlations in the system of gated metallic wires with the applied Zeeman magnetic field}
\author{V. Apinyan} 
\altaffiliation[e-mail:]{v.apinyan@intibs.pl}
\author{T. Kope\'c} 
\affiliation{W{\l}odzimierz Trzebiatowski Institute of Low Temperature and Structure Research, Polish Academy of Sciences\\ 50-422, ul. Ok{\'o}lna 2, Wroc\l{}aw, Poland 
}
%
\begin{abstract}
We have studied the electron-electron interactions in the system composed of two metallic wires, placed in the external magnetic and electric fields. The interactions between the electrons in the wires have been taken into account within the usual Hubbard model. We have considered both half-filling and partial-filling limits for the occupation of the atomic lattice sites. We show the existence of the excitonic pairing in this low-dimensional system and calculate the excitonic order parameter in different electron-electron interaction regime, magnetic field and temperature. We demonstrate that the usual Hubbard-$U$ interaction leads to strong electron localization which enhance the local antiferromagnetic order in the system. 
\end{abstract}     



\maketitle
\section{\label{sec:Section_1} Introduction}

The correlation effects in low-dimensional systems are more significant than in higher dimensions \cite{cite_1, cite_2, cite_3}. Their behavior is governed by many-body interactions, and the single-excitation picture breaks down \cite{cite_4}. As an example of such system is the one-dimensional (1D)quantum wire where the quantum coherence effects influence the electronic transport properties in such systems \cite{cite_5}, and the usual formula for the electrical resistance gets failed. Recently, it has been argued that the van der Waals interaction scheme between two separated metallic wires is qualitatively wrong \cite{cite_6}, which is especially the case of the electronic nanostructures that have zero energy gap. In general the binding energy of a material is determined by the spacial confinement of excitons \cite{cite_7, cite_8}. The study of excitons in low-dimensional materials is especially interesting because of the complicated many body physics that governs the formation of the excitons \cite{cite_9, cite_10}. The energy utilization in modern solar cells photovoltaics and optoelectronic devices based on the excitonic energy transfer establishes distinct differences with the traditional charge carrier transfer \cite{cite_10, cite_11, cite_12, cite_13}. Recently, the excitonic physics has emerged also in quantum information science \cite{cite_14, cite_15}. Being the neutral particles, the excitons can not move under the influence of the electric field potential, meanwhile the controlling of excitons could open new perspectives for the replacement of traditional transistors and enable a new way of data communication and processing within especially designed quantum circuit \cite{cite_16}.

In the present paper we investigated the excitons in the system composed of two metallic wires. For this purpose, a bi-wire Hubbard model is considered, and the effects of the external gate potential and magnetic fields are properly included. We show how this model is convenient for the full control of the excitonic properties in the system which include a very reach number of physical parameters that can be tuned experimentally. We show how the calculated values of the chemical potential affect the energy scales of different excitonic order parameters and we show that the large values of the Hubbard interaction energy enhance the antiferromagnetic order in the system and stabilize the charge density fluctuations in the system. The results in the paper are sound with respect to recent advances in the physics of excitons, and contain a large amount of information about the physical parameters which are important and helpful for further experimental investigations in the field. 

The paper is organized as follows: in the Section \ref{sec:Section_2} we introduce the bi-wire Hubbard Hamiltonian and we decouple the interacting terms. In the Section \ref{sec:Section_3}, we derive the system of self-consistent equation for the important physical quantities and we obtain the energy spectrum of bi-wire structure. The Section \ref{sec:Section_4} is devoted to the discussion of the obtained results and, finally, in Section \ref{sec:Section_5} we give a short conclusion to our paper.

\section{\label{sec:Section_2} The Hamiltonian of the system}
%
Our system is composed of two metallic wires. We write the bi-wire Hubbard Hamiltonian of the system in the form
\begin{eqnarray}
\hat{\cal{H}}&=&-t_0\sum_{\left\langle {\bf{r}}{\bf{r}}'\right\rangle,\ell=1,2,\sigma}\left(\hat{a}^{\dag}_{\ell\sigma}\left({\bf{r}}\right)\hat{a}_{\ell\sigma}\left({\bf{r}}'\right)+h.c.\right)
	\nonumber\\
	&&-t_1\sum_{{\bf{r}},\sigma}\left(\hat{a}^{\dag}_{1\sigma}\left({\bf{r}}\right)\hat{a}_{2\sigma}\left({\bf{r}}'\right)+h.c.\right)
	\nonumber\\
	&&+U\sum_{{\bf{r}},\ell}\hat{n}_{\ell\uparrow}\left({\bf{r}}\right)\hat{n}_{\ell\downarrow}\left({\bf{r}}\right)+W\sum_{{\bf{r}}\sigma\sigma'}\hat{n}_{1\sigma}({\bf{r}})\hat{n}_{2\sigma'}({\bf{r}})
	\nonumber\\
&& +\frac{V}{2}\sum_{{\bf{r}}}\left(\hat{n}_{2}\left({\bf{r}}\right)-\hat{n}_{1}\left({\bf{r}}\right)\right)-\mu\sum_{{\bf{r}}\ell}\hat{n}_{\ell}\left({\bf{r}}\right)
\nonumber\\
&&-g\mu_{\rm B}B\sum_{{\bf{r}},\ell}\left(\hat{n}_{\ell\uparrow}\left({\bf{r}}\right)-\hat{n}_{\ell\downarrow}\left({\bf{r}}\right)\right).
	\label{Equation_1}
\end{eqnarray}
%
%
\begin{figure}
	\includegraphics[scale=1.5]{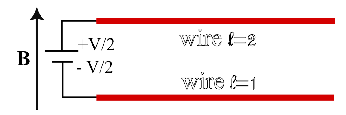}
	\caption{\label{fig:Fig_1}(Color online) The structure of the AB double-layer (DL) graphene in the external electric field potential $V$. The layers of the system have been indicated as $\ell=1$ (the bottom layer) and $\ell=2$ (the upper layer). In the picture, the $A$, $\tilde{A}$ atomic sites are represented by the black balls, and the $B$,$\tilde{B}$ atomic sites are represented by green balls.}
\end{figure} 
%
%
The first term describes the hopping of electrons in the wires. The operators $\hat{a}^{\dag}_{\ell\sigma}\left({\bf{r}}\right)$ and $\hat{a}_{\ell\sigma}\left({\bf{r}}\right)$ are the creation and annihilation operators for the electrons. The index $\ell$ denotes the wire ($\ell=1$ describes the bottom wire and $\ell=2$ the top wire) and $\sigma$ is the spin of the electrons which has two directions $\sigma=\uparrow,\downarrow$. The summation $\left\langle ... \right\rangle $ is over the nearest neighbor atomic sites in the wires and $t_0$ is the electron hopping amplitude which we put equal $t_0=1$ as the unit of measure of the energy scales. The parameter $t_1$, in the second term, is the electron hopping between the wires. Next, $U$, in the third term, is the on-site Hubbard interaction, which couples the electron density operators with opposite spin directions $\hat{n}_{\ell\uparrow}$ and $\hat{n}_{\ell\downarrow}$. They are defined as 
\begin{eqnarray}
\hat{n}_{\ell\sigma}=\hat{a}^{\dag}_{\ell\sigma}\left({{\bf{r}}}\right)\hat{a}_{\ell\sigma}\left({{\bf{r}}}\right).
\label{Equation_2}
\end{eqnarray}
Next, $\mu$, Eq.(\ref{Equation_1}), is the chemical potential, i.e., the minimum energy cost for arbitrary single-particle excitation in the system. We suppose that at the equilibrium the chemical potential is the same for two wires. The inter wire electron-electron interaction is given by Coulomb interaction energy term-$W$, in Eq.(\ref{Equation_1}). The applied external gate potential $V$ is included in the form of coupling with the electron densities in the wires and this term describes the interaction of the electron gas with the electric field potential (see in Fig.~\ref{fig:Fig_1}). The electron densities $\hat{n}_{1}\left({{\bf{r}}}\right)$ and $\hat{n}_{2}\left({{\bf{r}}}\right)$ are defined as
\begin{eqnarray}
	\hat{n}_{\ell}\left({{\bf{r}}}\right)=\sum_{\sigma}\hat{a}^{\dag}_{\ell\sigma}\left({{\bf{r}}}\right)\hat{a}_{\ell\sigma}\left({{\bf{r}}}\right).
	\label{Equation_3}
\end{eqnarray}
The Zeeman effect of the static magnetic field, applied to the system, is given by the last term in Eq.(\ref{Equation_1}), where $g$ is the Land{\'e} $g$-factor and $\mu_{\rm B}$ is the Bohr magneton (we chosen the units where $\hbar=1$ and $\mu_{B}=1$).  

The density-density product $\hat{n}_{\ell\uparrow}\left({\bf{r}}\right)\hat{n}_{\ell\downarrow}\left({\bf{r}}\right)$ in the following form
\begin{eqnarray}
\hat{n}_{\ell\uparrow}\left({\bf{r}}\right)\hat{n}_{\ell\downarrow}=\frac{1}{4}\hat{n}^{2}_{\ell\uparrow}\left({\bf{r}}\right)-\hat{S}^{2}_{\ell \rm{z}}\left({\bf{r}}\right),
	\label{Equation_4}
\end{eqnarray}
where $\hat{n}^{2}_{\ell\uparrow}\left({\bf{r}}\right)$ is given in Eq.(\ref{Equation_3}) and $S^{2}_{z}\left({\bf{r}}\right)$ is the $z$-component of the generalized spin operator, defined as
\begin{eqnarray}
	\hat{S}_{\ell \rm{z}}\left({\bf{r}}\right)=\frac{1}{2}\sum_{\alpha,\beta=\uparrow\downarrow}\hat{a}^{\dag}_{\ell\alpha}\left({\bf{r}}\right)\hat{\sigma}_{z\alpha\beta}\hat{a}_{\ell\beta}\left({\bf{r}}\right)
	\nonumber\\
	=\frac{1}{2}\left(\hat{a}^{\dag}_{\ell\uparrow}\left({\bf{r}}\right)\hat{a}_{\ell\uparrow}\left({\bf{r}}\right)-\hat{a}^{\dag}_{\ell\downarrow}\left({\bf{r}}\right)\hat{a}_{\ell\downarrow}\left({\bf{r}}\right)\right)
	\label{Equation_5}
\end{eqnarray}
and $\hat{\sigma}_{z}$ is the $\rm{z}$ component of the Pauli matrix vector. And Hubbard-$U$ term in the Eq.(\ref{Equation_1}) could be rewritten as
\begin{eqnarray}
\hat{\cal{H}}_{U}=U\sum_{{\bf{r}}}\left[\frac{1}{4} \hat{n}^{2}_{\ell\uparrow}\left({\bf{r}}\right)-\hat{S}^{2}_{\ell \rm{z}}\left({\bf{r}}\right)\right].
\label{Equation_6}
\end{eqnarray}	
Next, we write, for the convenience, the $W$-interaction term in the following form
\begin{eqnarray}
&&\hat{\cal{H}}_{W}=W\sum_{{\bf{r}}\sigma\sigma'}\hat{n}_{1\sigma}({\bf{r}})\hat{n}_{2\sigma'}({\bf{r}})=
\nonumber\\
&&=W\sum_{{\bf{r}}\sigma\sigma'}\hat{a}^{\dag}_{1\sigma}\left({\bf{r}}\right)\hat{a}_{1\sigma}\left({\bf{r}}'\right)\hat{a}^{\dag}_{2\sigma'}\left({\bf{r}}\right)\hat{a}_{2\sigma'}\left({\bf{r}}\right)
\nonumber\\
&&=2W\sum_{{\bf{r}}\sigma}\hat{n}_{1\sigma}\left({\bf{r}}\right)-W\sum_{{\bf{r}}\sigma\sigma'}|\hat{\xi}_{\sigma\sigma'}\left({\bf{r}}\right)|^{2},
\label{Equation_7}
\end{eqnarray}
where we have introduced the following operators
\begin{eqnarray}
{\hat{\xi}}^{\dag}_{\sigma\sigma'}\left({\bf{r}}\right)=\hat{a}^{\dag}_{1\sigma}\left({\bf{r}}\right)\hat{a}_{2\sigma'}\left({\bf{r}}\right)
\nonumber\\
{\hat{\xi}}_{\sigma\sigma'}\left({\bf{r}}\right)=\hat{a}^{\dag}_{2\sigma'}\left({\bf{r}}\right)\hat{a}_{1\sigma}\left({\bf{r}}\right).
\label{Equation_8}
\end{eqnarray}
This form of the $W$-interaction, in Eq.(\ref{Equation_7}), is more convenient for further decoupling procedure, described in the next section. 
%
\section{\label{sec:Section_3} Self-consistent equations}
%
\subsection{\label{sec:Section_3_1} Hubbard-Stratonovich decoupling}
%
Furthermore, we pass into the Grassmann complex variable representation \cite{cite_17}, in which we replace the fermionic operators with the complex numbers. Then, the action of the system of two wires could be written as
\begin{eqnarray}
{\cal{S}}\left[\bar{a}_{1\sigma},a_{1\sigma},\bar{a}_{2\sigma}, a_{2\sigma}\right]=\int^{\beta}_{0}d\tau {\hat{\cal{H}}}\left(\tau\right)
\nonumber\\
+{\cal{S}}_{{\rm B}}\left[\bar{a}_{1\sigma},a_{1\sigma},\bar{a}_{2\sigma}, a_{2\sigma}\right],
\label{Equation_9}
\end{eqnarray}
where $\beta$, in Eq.(\ref{Equation_6}), is $\beta=1/k_{B}T$, with $T$ being the temperature of the system. The integration variable $\tau$ is the Matsubara imaginary time and ${\cal{S}}_{{\rm B}}$ is the fermionic Berry term. This last term is defined as
\begin{eqnarray}
{\cal{S}}_{{\rm B}}\left[\bar{a}_{1\sigma},a_{1\sigma},\bar{a}_{2\sigma}, a_{2\sigma}\right]=\sum_{\ell=1,2}\int^{\beta}_{0}d\tau \bar{a}_{\ell\sigma}\left({\bf{r}}\right)\partial_{\tau}a_{\ell\sigma}\left({\bf{r}}\right).
\nonumber\\
\label{Equation_10}
\end{eqnarray}
The partition function of the system, in the path integral representation is
\begin{eqnarray}
{\cal{Z}}=\int\left[{\cal{D}}\bar{a}_{1}{\cal{D}}{a}_{1}\right]\left[{\cal{D}}\bar{a}_{2}{\cal{D}}{a}_{2}\right]e^{-	{\cal{S}}\left[\bar{a}_{1\sigma},a_{1\sigma},\bar{a}_{2\sigma}, a_{2\sigma}\right]}.	
\label{Equation_11}
\end{eqnarray}
Now, we can decouple the first biquadratic term in Eq.(\ref{Equation_6}) via Hubbard-Stratonovich transformation rule, i.e.,
\begin{eqnarray}
&&e^{-\int^{\beta}_{0}d\tau\frac{U}{4}\sum_{{\bf{r}}\ell}{n}^{2}_{\ell}\left({\bf{r}}\tau\right)}
\nonumber\\
&&=\int\left[{\cal{D}}V_{\ell}\right]e^{\int^{\beta}_{0}d\tau\sum_{{\bf{r}}\ell}\left[-\frac{V^{2}_{\ell}\left({\bf{r}}\tau\right)}{U}+i{n}_{\ell}\left({\bf{r}}\tau\right)V_{\ell}\left({\bf{r}}\tau\right)\right]}.
\label{Equation_12}
\end{eqnarray}
Then, we calculate the saddle-point value $V_{\rm 0 \ell}$ of the decoupling field $V_{\ell}\left({\bf{r}}\tau\right)$. We get
\begin{eqnarray}
V_{0\ell}=\frac{iU}{2}\bar{n}_{\ell},
\label{Equation_13}
\end{eqnarray}
where $\bar{n}_{\ell}$ means the statistical average of the fermion density
\begin{eqnarray}
&&\bar{n}_{\ell}=\left\langle ... \right\rangle=
\nonumber\\
&&=\frac{1}{{\cal{Z}}}\int\int\left[{\cal{D}}\bar{a}_{1}{\cal{D}}{a}_{1}\right]\left[{\cal{D}}\bar{a}_{2}{\cal{D}}{a}_{2}\right]...e^{-	{\cal{S}}\left[\bar{a}_{1\sigma},a_{1\sigma},\bar{a}_{2\sigma}, a_{2\sigma}\right]}.
\nonumber\\	
\label{Equation_14}
\end{eqnarray}
Then, the contribution to the action, coming from the decoupling field, is 
\begin{eqnarray}
{\cal{S}}\left[V_{0\ell}\right]=\sum_{{\bf{r}}}\int^{\beta}_{0}d\tau \frac{U}{2}\bar{n}_{\ell}n_{\ell}\left({\bf{r}}\tau\right).
\label{Equation_15}
\end{eqnarray}
For the second term, in Eq.(\ref{Equation_6}), we have
\begin{eqnarray}
&&e^{\int^{\beta}_{0}d\tau{U}\sum_{{\bf{r}}\ell}S^{2}_{{\rm z}\ell}\left({\bf{r}}\tau\right)}=
\nonumber\\
&&=\int\left[{\cal{D}}\zeta_{\ell}\right]e^{\int^{\beta}_{0}d\tau\sum_{{\bf{r}}\ell}\left[-\frac{{\zeta}^{2}_{\ell}\left({\bf{r}}\tau\right)}{U}+{\zeta}_{\ell}\left({\bf{r}}\tau\right)S_{{\rm z}\ell}\left({\bf{r}}\tau\right)\right]}
\label{Equation_16}
\end{eqnarray}
and we get the saddle-point value ${\zeta}_{0}\left({\bf{r}}\tau\right)$ for the decoupling field ${\zeta}_{\ell}\left({\bf{r}}\tau\right)$
\begin{eqnarray}
{\zeta}_{0\ell}=U\bar{S}_{{\rm z}\ell}.
\label{Equation_17}
\end{eqnarray}
Furthermore, the contribution to the action reads as
\begin{eqnarray}
{\cal{S}}\left[{\zeta}_{0\ell}\right]=-2{U}\sum_{{\bf{r}}\ell}\int^{\beta}_{0}d\tau S_{{\rm z} \ell}\left({\bf{r}}\tau\right)\bar{S}_{{\rm z} \ell}.
\label{Equation_18}
\end{eqnarray}
The same type of Hubbard-Stratonovich transformation could be written also for the second term in Eq.(\ref{Equation_7}). We have
\begin{eqnarray}
&&e^{W\int^{\beta}_{0}d\tau\sum_{{\bf{r}}\sigma\sigma'}|\hat{\xi}_{\sigma\sigma'}\left({\bf{r}}\tau\right)|^{2}}
\nonumber\\
&&=\int\left[{\cal{D}}\bar{\Lambda}{\cal{D}}\Lambda\right]e^{\sum_{{\bf{r}}\sigma\sigma'}\int^{\beta}_{0}d\tau-\frac{1}{W}|\Lambda_{\sigma\sigma'}\left({\bf{r}}\tau\right)|^{2}}\times
\nonumber\\
&&\times\exp{\left({\xi}_{\sigma\sigma'}\left({\bf{r}}\tau\right)\bar{\Lambda}_{\sigma\sigma'}\left({\bf{r}}\tau\right)+\bar{\Lambda}_{\sigma\sigma'}\left({\bf{r}}\tau\right)\bar{\xi}_{\sigma\sigma'}\left({\bf{r}}\tau\right)\right)}.
\nonumber\\
\label{Equation_19}
\end{eqnarray}
The functional derivation with respect to ${\Lambda}_{\sigma\sigma'}\left({\bf{r}}\tau\right)$ gives us the saddle-point value for $\bar{\Lambda}_{\sigma\sigma'}\left({\bf{r}}\tau\right)$, i.e.,
\begin{eqnarray}
\bar{\Lambda}_{0\sigma\sigma'}=W\left\langle \hat{\xi}_{\sigma\sigma'}\left({\bf{r}}\tau\right) \right\rangle=W\left\langle \bar{a}_{1\sigma}\left({\bf{r}}\tau\right)a_{2\sigma'}\left({\bf{r}}\tau\right)\right\rangle.
\label{Equation_20}
\end{eqnarray}
Furthermore, the functional derivation with respect to $\bar{\Lambda}_{\sigma\sigma'}\left({\bf{r}}\tau\right)$ gives us the saddle-point value of ${\Lambda}_{\sigma\sigma'}\left({\bf{r}}\tau\right)$, i.e.,
\begin{eqnarray}
{\Lambda}_{0\sigma\sigma'}=W\left\langle {\xi}_{\sigma\sigma'}\left({\bf{r}}\tau\right) \right\rangle=W\left\langle \bar{a}_{2\sigma'}\left({\bf{r}}\tau\right)a_{1\sigma}\left({\bf{r}}\tau\right)\right\rangle.
\label{Equation_21}
\end{eqnarray}
In fact, the saddle-point values $\bar{\Lambda}_{0\sigma\sigma'}$ and ${\Lambda}_{0\sigma\sigma'}$ are the subject of the excitonic gap parameters ${\bar{\Delta}}_{\sigma\sigma'}$ and complex conjugate $\Delta_{\sigma\sigma'}$ 
Then, the contribution to the action, coming from these two saddle-point values is
\begin{eqnarray}
{\cal{S}}_{{\rm W}}=-\sum_{{\bf{r}}\sigma\sigma'}\int^{\beta}_{0}d\tau {\Delta}_{\sigma\sigma'}\bar{a}_{2\sigma'}\left({\bf{r}}\tau\right){a}_{1\sigma'}\left({\bf{r}}\tau\right)
\nonumber\\
-\sum_{{\bf{r}}\sigma\sigma'}\int^{\beta}_{0}d\tau \bar{\Delta}_{\sigma\sigma'}\bar{a}_{1\sigma}\left({\bf{r}}\tau\right){a}_{2\sigma}\left({\bf{r}}\tau\right).
\label{Equation_22}
\end{eqnarray}
In the next section, we derive the form of Green's function matrix and calculate the band structure in the system of two metallic wires. 
%
\subsection{\label{sec:Section_3_2} The action and energy spectrum}
%
The fermionic action in Eq.(\ref{Equation_9}) could be rewritten in the Fourier space representation. Taking into account the actions in Eqs.(\ref{Equation_15}), (\ref{Equation_18}) and (\ref{Equation_22}) we can write
\begin{eqnarray}
&&{\cal{S}}\left[\bar{a}_{1}, a_{1}, \bar{a}_{2}, a_{2}\right]=-\frac{1}{\beta{N}}\sum_{{\bf{k}}\nu_{n}}\sum_{\sigma}\left[\mu-2W+\left(-1\right)^{\sigma}\times \right.
\nonumber\\
&&\left.\times{g\mu_{B}B}+i\nu_{n}+\frac{V}{2}-\frac{U\bar{n}_{1}}{2}+\left(-1\right)^{\sigma}\Delta^{\left(1\right)}_{{\rm AFM}}\right.
\nonumber\\
&&\left.+4t_0\cos\left(ka\right)\right]\bar{a}_{1\sigma}\left({\bf{k}},\nu_{n}\right){a}_{1\sigma}\left({\bf{k}},\nu_{n}\right)
\nonumber\\
&&-\frac{1}{\beta{N}}\sum_{{\bf{k}}\nu_{n}}\sum_{\sigma}\left[\mu-2W+\left(-1\right)^{\sigma}g\mu_{B}B+i\nu_{n}-\frac{V}{2}\right.
\nonumber\\
&&\left.-\frac{U\bar{n}_{1}}{2}+\left(-1\right)^{\sigma}\Delta^{\left(2\right)}_{{\rm AFM}}+4t_0\cos\left(ka\right)\right]\times
\nonumber\\
&&\times{\bar{a}_{2\sigma}\left({\bf{k}},\nu_{n}\right){a}_{2\sigma}\left({\bf{k}},\nu_{n}\right)}
\nonumber\\
&&-\frac{t_1+\bar{\Delta}}{\beta{N}}\sum_{{\bf{k}},\nu_{n}}\sum_{\sigma}\bar{a}_{1\sigma}\left({\bf{k}},\nu_{n}\right){a}_{2\sigma}\left({\bf{k}},\nu_{n}\right)
\nonumber\\
&&-\frac{t_1+{\Delta}}{\beta{N}}\sum_{{\bf{k}},\nu_{n}}\sum_{\sigma}\bar{a}_{1\sigma}\left({\bf{k}},\nu_{n}\right){a}_{2\sigma}\left({\bf{k}},\nu_{n}\right).
\label{Equation_23}
\end{eqnarray}
We introduce here the composite Nambu spinors for our problem
\begin{eqnarray} 
\bar{\psi}({\bf{k}}\nu_{n})=\left(\bar{a}_{1\uparrow}({\bf{k}}\nu_{n}),\bar{a}_{1\downarrow}({\bf{k}},\nu_{n}),\bar{b}_{1\uparrow}({\bf{k}},\nu_{n}),\bar{b}_{2\downarrow}({\bf{k}},\nu_{n})\right)
	\nonumber\\
	\label{Equation_24}
\end{eqnarray}
and
\begin{eqnarray} 
\footnotesize
\arraycolsep=0pt
\medmuskip = 0mu
{\psi}({\bf{k}},\nu_{n})=\left(
\begin{array}{c}
{a}_{1\uparrow}({\bf{k}},\nu_{n})\\\\
{a}_{1\downarrow}({\bf{k}},\nu_{n})\\\\
{b}_{1\uparrow}({\bf{k}},\nu_{n}) \\\\
{b}_{2\downarrow}({\bf{k}},\nu_{n})
\end{array}
\right).
\label{Equation_25}
\end{eqnarray}
Next, the action of the system of two wires could be written in the following compact form
\begin{eqnarray} 
{\cal{S}}\left[\bar{\psi},\psi\right]=\frac{1}{\beta{N}}\sum_{{\bf{k}},\nu_{n}}\bar{\psi}({\bf{k}},\nu_{n}){\cal{G}}^{-1}\left({\bf{k}},\nu_{n}\right){\psi}({\bf{k}},\nu_{n}),
\nonumber\\
\label{Equation_26}
\end{eqnarray}
where we have introduced Gorkov matrix ${\cal{G}}^{-1}\left({\bf{k}},\nu_{n}\right)$
\begin{widetext}
\begin{eqnarray}
&&{\cal{G}}^{-1}({\bf{k}},\nu_{n})=\left(
\begin{array}{ccccrrrr}
-i\nu_{n}-\mu_{1\sigma} & 0 & -\bar{\Delta}_{\uparrow}-t_1 & 0\\
			0 & -i\nu_{n}-\mu_{2\sigma} & 0 &  -\bar{\Delta}_{\downarrow}-t_1\\
			-{\Delta}_{\uparrow}-{\gamma}_1 & 0 & -i\nu_{n}-\mu_{3\sigma} & 0 \\
			0 & -{\Delta}_{\downarrow}-{\gamma}_1 & 0 & -i\nu_{n}-\mu_{4\sigma}
\end{array}
\right).
\label{Equation_27}
\end{eqnarray}
\end{widetext}
We have introduced in Eq.(\ref{Equation_27}) the following effective chemical potentials 
\begin{eqnarray} 
\mu_{1}&=&\mu-2W+\frac{V}{2}-\frac{U}{4}\left(\frac{1}{\kappa}-\delta{\bar{n}}\right)+4t_0\cos\left(ka\right)
\nonumber\\
&&+g\mu_{B}B+\Delta^{\left(1\right)}_{{\rm AFM}},
\nonumber\\
\mu_{2}&=&\mu-2W+\frac{V}{2}-\frac{U}{4}\left(\frac{1}{\kappa}-\delta{\bar{n}}\right)+4t_0\cos\left(ka\right)
\nonumber\\
&&-g\mu_{B}B-\Delta^{\left(1\right)}_{{\rm AFM}},
\nonumber\\
\mu_{3}&=&\mu-\frac{V}{2}-\frac{U}{4}\left(\frac{1}{\kappa}+\delta{\bar{n}}\right)+4t_0\cos\left(ka\right)
\nonumber\\
&&+g\mu_{B}B+\Delta^{\left(2\right)}_{{\rm AFM}},
\nonumber\\
\mu_{4}&=&\mu-\frac{V}{2}-\frac{U}{4}\left(\frac{1}{\kappa}+\delta{\bar{n}}\right)+4t_0\cos\left(ka\right)
\nonumber\\
&&-g\mu_{B}B-\Delta^{\left(2\right)}_{{\rm AFM}}.
\label{Equation_28}
\end{eqnarray}
We suppose the staggered form of the antiferromagnetic order parameter between the wires, i.e., $\Delta^{\left(1\right)}_{{\rm AFM}}=-\Delta^{\left(2\right)}_{{\rm AFM}}$. The energy spectrum of the problem can be recovered from the equation $\det\left[{\cal{G}}^{-1}({\bf{k}},\nu_{n})\right]=0$. We get the band-structure in the form
\begin{eqnarray} 
\varepsilon_{1}\left({\bf{k}}\right)=0.5\left[-\mu_{1}-\mu_{3}-\sqrt{\left(\mu_1-\mu_3\right)^{2}+4|\Delta_{\uparrow}+t_1|^{2}}\right],
\nonumber\\
\varepsilon_{2}\left({\bf{k}}\right)=0.5\left[-\mu_{1}-\mu_{3}+\sqrt{\left(\mu_1-\mu_3\right)^{2}+4|\Delta_{\uparrow}+t_1|^{2}}\right],
\nonumber\\
\varepsilon_{3}\left({\bf{k}}\right)=0.5\left[-\mu_{2}-\mu_{4}-\sqrt{\left(\mu_2-\mu_4\right)^{2}+4|\Delta_{\downarrow}+t_1|^{2}}\right],
\nonumber\\
\varepsilon_{4}\left({\bf{k}}\right)=0.5\left[-\mu_{2}-\mu_{4}+\sqrt{\left(\mu_2-\mu_4\right)^{2}+4|\Delta_{\downarrow}+t_1|^{2}}\right].
\nonumber\\
\label{Equation_29}
\end{eqnarray}
We see, in Eq.(\ref{Equation_29}), that the energies $\varepsilon_{1}\left({\bf{k}}\right)$ and $\varepsilon_{2}\left({\bf{k}}\right)$ enunciate the correspondence to the spin direction $\sigma=\uparrow$ and the energies $\varepsilon_{3}\left({\bf{k}}\right)$ and $\varepsilon_{4}\left({\bf{k}}\right)$ correspond to the spin direction $\sigma=\downarrow$.  
Furthermore, in the next section, we construct the system of self-consistent equations for the considered problem. 
%
\subsection{\label{sec:Section_3_3} The system of equations}
%
We suppose here different values for the average number of particle occupation at the atomic sites positions, in both metallic wires. Therefore, we write
\begin{eqnarray}
\bar{n}_{1}+\bar{n}_{2}=\frac{1}{\kappa}.
\label{Equation_30}
\end{eqnarray}
Particularly, the value $\kappa=0.5$ corresponds to the case of half-filling. In order to study the average charge redistribution in the system we consider the function $\delta{\bar{n}}=\bar{n}_{2}-\bar{n}_{1}$, which is, indeed, the average charge imbalance between wires. For the homogeneous case, we have $\bar{n}_{\ell\uparrow}=\bar{n}_{\ell\downarrow}$ (with $\ell=1,2$). From the other hand, it is clear that $\bar{n}_{1}\neq\bar{n}_{2}$. Thus, we can write $\bar{n}_{1\uparrow}+\bar{n}_{2\uparrow}=\bar{n}_{1\downarrow}+\bar{n}_{2\downarrow}$.
Then, the system of coupled equations could be written as
\begin{eqnarray}
&&\bar{n}_{1\uparrow}+\bar{n}_{2\uparrow}=\frac{1}{2\kappa},
\nonumber\\
&&\bar{n}_{2\uparrow}-\bar{n}_{1\downarrow}=\frac{\delta{\bar{n}}}{2},
\nonumber\\
&&\Delta_{\uparrow}=W\left\langle \bar{a}_{1\uparrow}a_{2\uparrow}\right\rangle,
\nonumber\\
&&\Delta_{\downarrow}=W\left\langle \bar{a}_{1\downarrow}a_{2\downarrow}\right\rangle,
\nonumber\\
&&\Delta_{{\rm AFM}}=\frac{U}{2}\left(\bar{n}_{1\uparrow}-\bar{n}_{1\downarrow}\right).
\label{Equation_31}
\end{eqnarray}
Furthermore, after summing over the fermionic Matsubara frequencies $\nu_{n}$ we get
\begin{eqnarray}
&&\frac{1}{N}\sum^{4}_{i=1}\sum_{{\bf{k}}}\alpha_{i}\left({\bf{k}}\right)n_{F}\left(\varepsilon_{i}\left({\bf{k}}\right)\right)=-\frac{1}{2\kappa},
\nonumber\\
&&\frac{1}{N}\sum^{4}_{i=1}\sum_{{\bf{k}}}\beta_{i}\left({\bf{k}}\right)n_{F}\left(\varepsilon_{i}\left({\bf{k}}\right)\right)=-\frac{\delta{\bar{n}}}{2},
\nonumber\\
&&W\frac{\Delta_{\uparrow}+t_1}{N}\sum^{4}_{i=1}\sum_{{\bf{k}}}\gamma^{\left(1\right)}_{i}\left({\bf{k}}\right)n_{F}\left(\varepsilon_{i}\left({\bf{k}}\right)\right)=-\Delta_{\uparrow},
\nonumber\\
&&W\frac{\Delta_{\downarrow}+t_1}{N}\sum^{4}_{i=1}\sum_{{\bf{k}}}\gamma^{\left(2\right)}_{i}\left({\bf{k}}\right)n_{F}\left(\varepsilon_{i}\left({\bf{k}}\right)\right)=-\Delta_{\downarrow},
\nonumber\\
&&\frac{U}{2N}\sum^{4}_{i=1}\sum_{{\bf{k}}}\delta_{i}\left({\bf{k}}\right)n_{F}\left(\varepsilon_{i}\left({\bf{k}}\right)\right)=-\Delta_{{\rm AFM}},
\label{Equation_32}
\end{eqnarray}
where 
\begin{eqnarray}
n_{F}\left(x\right)=\frac{1}{e^{x-\mu}+1},
\label{Equation_33}
\end{eqnarray}
on the left-hand sides in Eq.(\ref{Equation_32}), is Fermi-Dirac distribution function. 
The coefficients $\alpha_{i}\left({\bf{k}}\right)$, $\beta_{i}\left({\bf{k}}\right)$, $\gamma^{\left(1\right)}_{i}\left({\bf{k}}\right)$, $\gamma^{\left(2\right)}_{i}\left({\bf{k}}\right)$ and $\delta_{i}\left({\bf{k}}\right)$ are given in Appendix \ref{sec:Section_5}.
%
\section{\label{sec:Section_4} Results and discussions}
\begin{figure}
	\includegraphics[width=8.5cm, height=16cm]{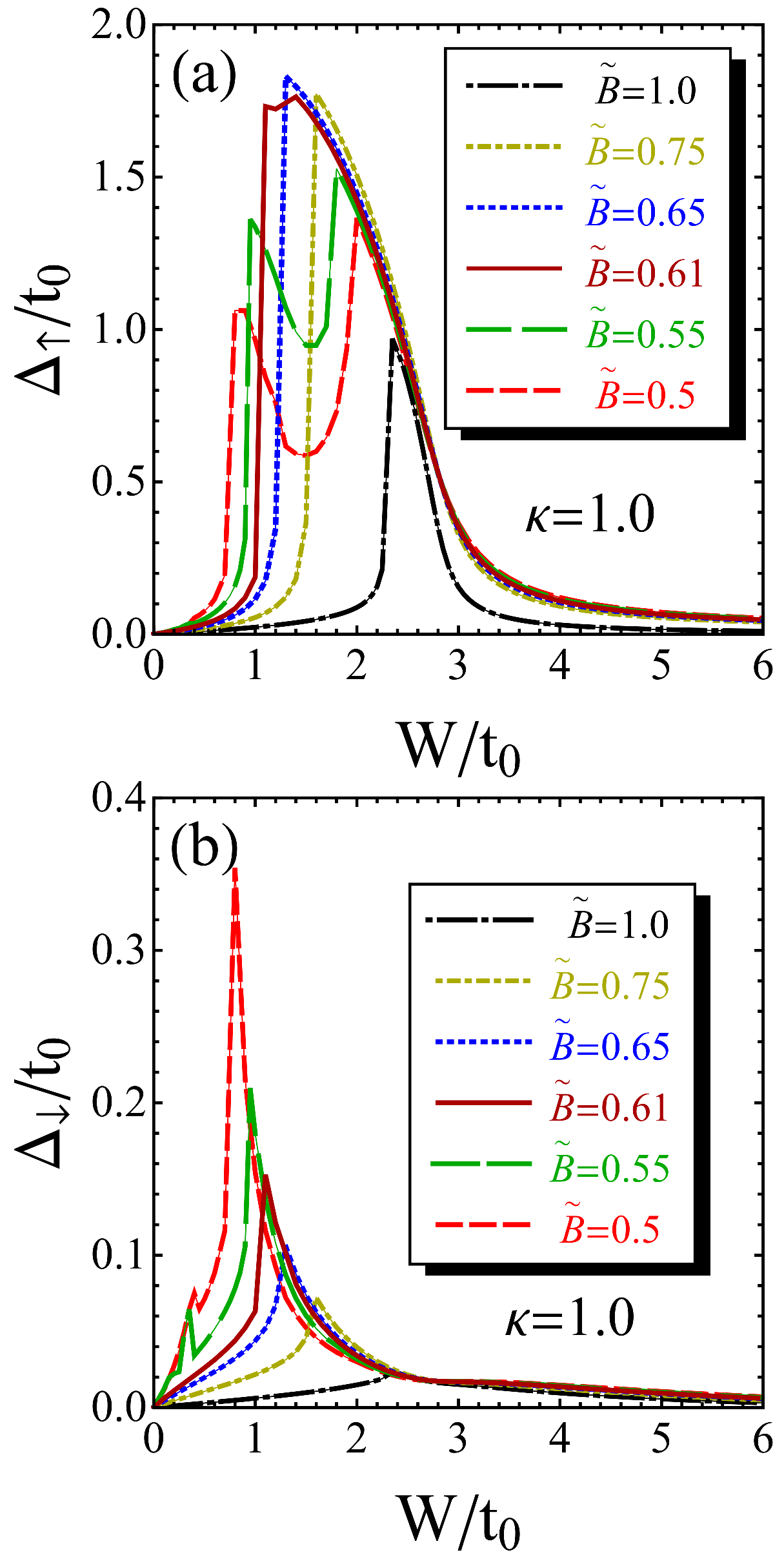}
	\caption{\label{fig:Fig_2}(Color online) The excitonic order parameters $\Delta_{\uparrow}$ (see panel (a)) and $\Delta_{\downarrow}$ (see panel (b)), calculated in Eq.(\ref{Equation_32}), as a function of the inter-wire Coulomb interaction potential $W$. The case of half-filling with $\kappa=1.0$ has been considered for different values of the external magnetic field parameter $\tilde{B}=\mu_{\rm B}B/t_{0}$. The external gate potential and local Hubbard-$U$ potential have been fixed at the values $V=t_{0}$ and $U=0.6t_{0}$. The calculations have been performed at the zero temperature limit $T=0.0$. The calculations have been performed at the zero temperature limit $T=0$.}
\end{figure}

%
In Fig.~\ref{fig:Fig_2}, we calculated the excitonic order parameters for different spin configurations $\Delta_{\uparrow}$ (see panel (a), in Fig.~\ref{fig:Fig_2}) and $\Delta_{\downarrow}$ (see panel (b), in Fig.~\ref{fig:Fig_2}), after Eq.(\ref{Equation_32}), given in Section \ref{sec:Section_3_3}, as a function of the inter-wire Coulomb interaction potential $W$. The case of half-filling have been considered when the inverse of average number of particles at the given lattice site positions is fractional $\kappa=1/n_{\rm fill}=1.0$ and $n_{\rm fill}$ is the number of particles at the individual lattice site position. For example, $\kappa=0.5$ corresponds the case of half-filling, when the maximum number of particles at the lattice sites is one, i.e., $n_{\rm fill}=1.0$. 
%
%
hopping amplitude $t_0$.
%
%
Different values of the external magnetic field parameter $\tilde{B}=\mu_{\rm B}B/t_{0}$ have been considered at the fixed values of the electric field gate potential $V$ and Hubbard-$U$ interaction parameter: $V=t_{0}$ and $U=0.6t_{0}$. We observe that when lowering the magnetic field parameter, the excitonic excitonic peaks displacing to the left on the $W$-axis, while there a critical value of the magnetic field $B_{\rm C}$ at which the excitonic transition lines, corresponding to $\Delta_{\sigma}$, start to split into two-peak like curves. For $B<B_{\rm C}$ the two-peak like structure remains and is more pronounced for $\Delta_{\uparrow}$. Moreover, an interesting behavior is visible for different limits of the magnetic field. Namely, for $\tilde{B}>\tilde{B}_{\rm C}$ the magnitude of the excitonic order parameter $\Delta_{\uparrow}$ decreases when increasing $\tilde{B}$ up to value $\tilde{B}=1.0$ and in the transition region $\tilde{B}<\tilde{B}_{\rm C}$ the excitonic order parameter $\Delta_{\uparrow}$ increases when increasing the magnetic field parameter $\tilde{B}$. Those opposite behaviors are the results of the critical value $B_{\rm C}$. Concomitantly, the order parameter $\Delta_{\downarrow}$ (see panel (b), in Fig.~\ref{fig:Fig_2}) increases continuously when decreasing $\tilde{B}$, and the existence of the critical value $B_{\rm C}$ doesn't affect this behavior. We observe also that the order parameter $\Delta_{\downarrow}$ is much smaller (of about one order of magnitude)that the excitonic order parameter $\Delta_{\uparrow}$.         

\begin{figure}
	\includegraphics[width=8.2cm, height=5.2cm]{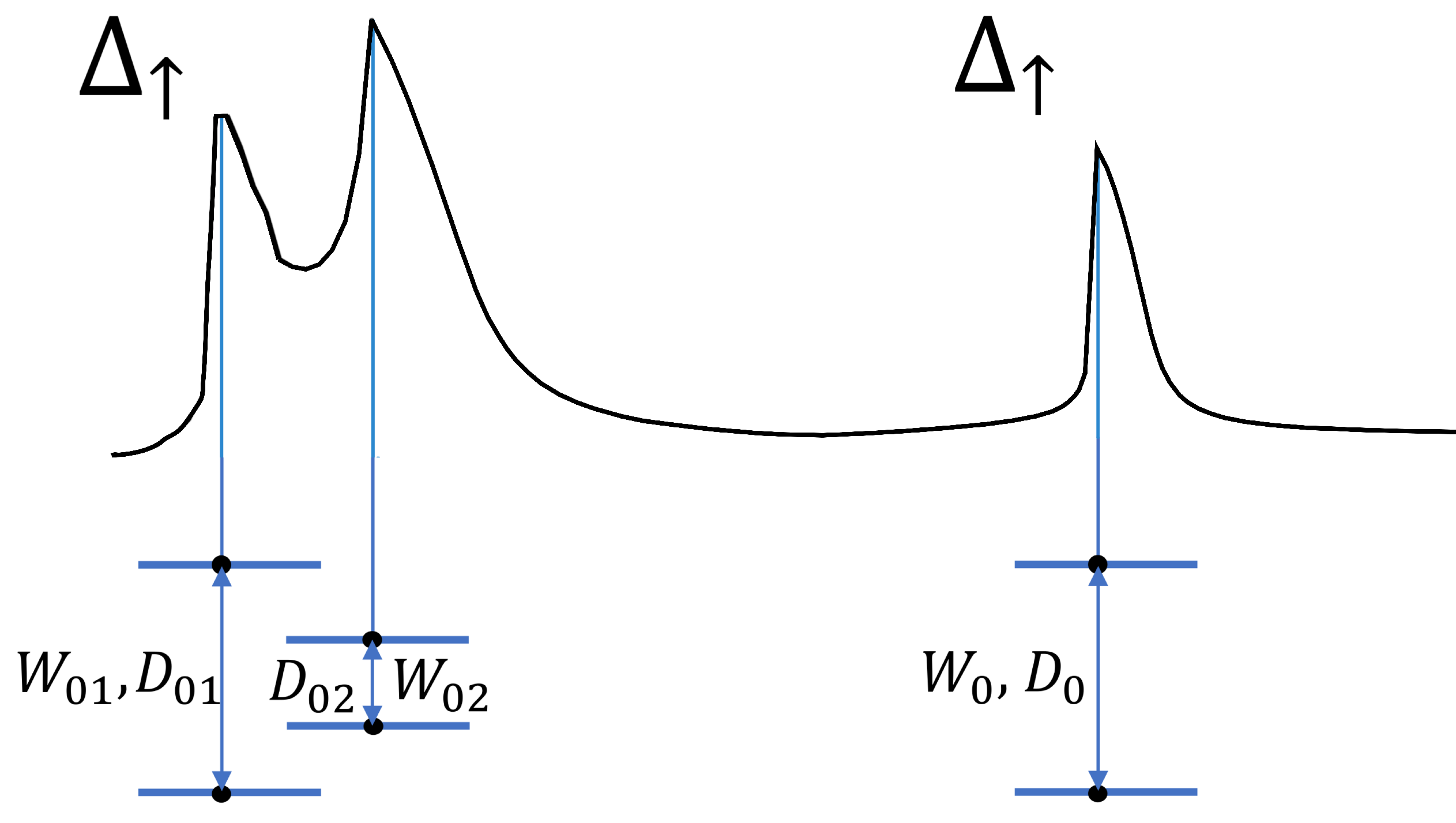}
	\caption{\label{fig:Fig_3}(Color online) (left picture) The excitonic order parameter $\Delta_{\uparrow}$ at $\tilde{B}<\tilde{B}_{\rm C}$. Two strong excitonic peaks arise when varying the inter-chain interaction potential $W$ or when changing the inter-chain separation distance. (right picture) The excitonic order parameter $\Delta_{\uparrow}$ at $\tilde{B}>\tilde{B}_{\rm C}$. The calculations have been performed at the zero temperature limit $T=0$.}
\end{figure}

\begin{figure}
	\includegraphics[width=8.5cm, height=20cm]{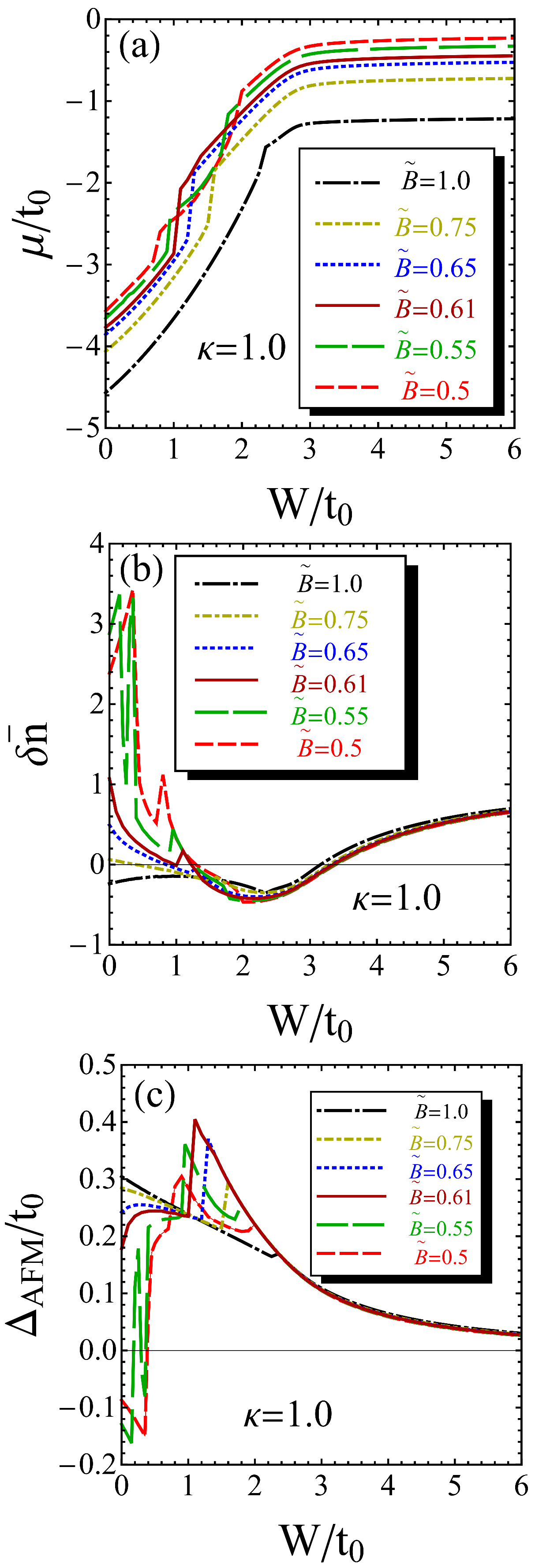}
	\caption{\label{fig:Fig_4}(Color online) The chemical potential (see panel (a)), average charge density imbalance between wires (see panel (b)) and antiferromagnetic order parameter $\Delta_{\rm AFM}$, calculated in Eq.(\ref{Equation_32}), as a function of the inter-wires Coulomb interaction potential $W$. The case of half-filling with $\kappa=1.0$ has been considered, for different values of the external magnetic field parameter $\tilde{B}=\mu_{\rm B}B/t_{0}$. The external gate potential and local Hubbard-$U$ potential have been fixed at the values $V=t_{0}$ and $U=0.6t_{0}$. The calculations have been performed at the zero temperature limit $T=0$.}
\end{figure}

\begin{figure}
	\includegraphics[scale=0.21]{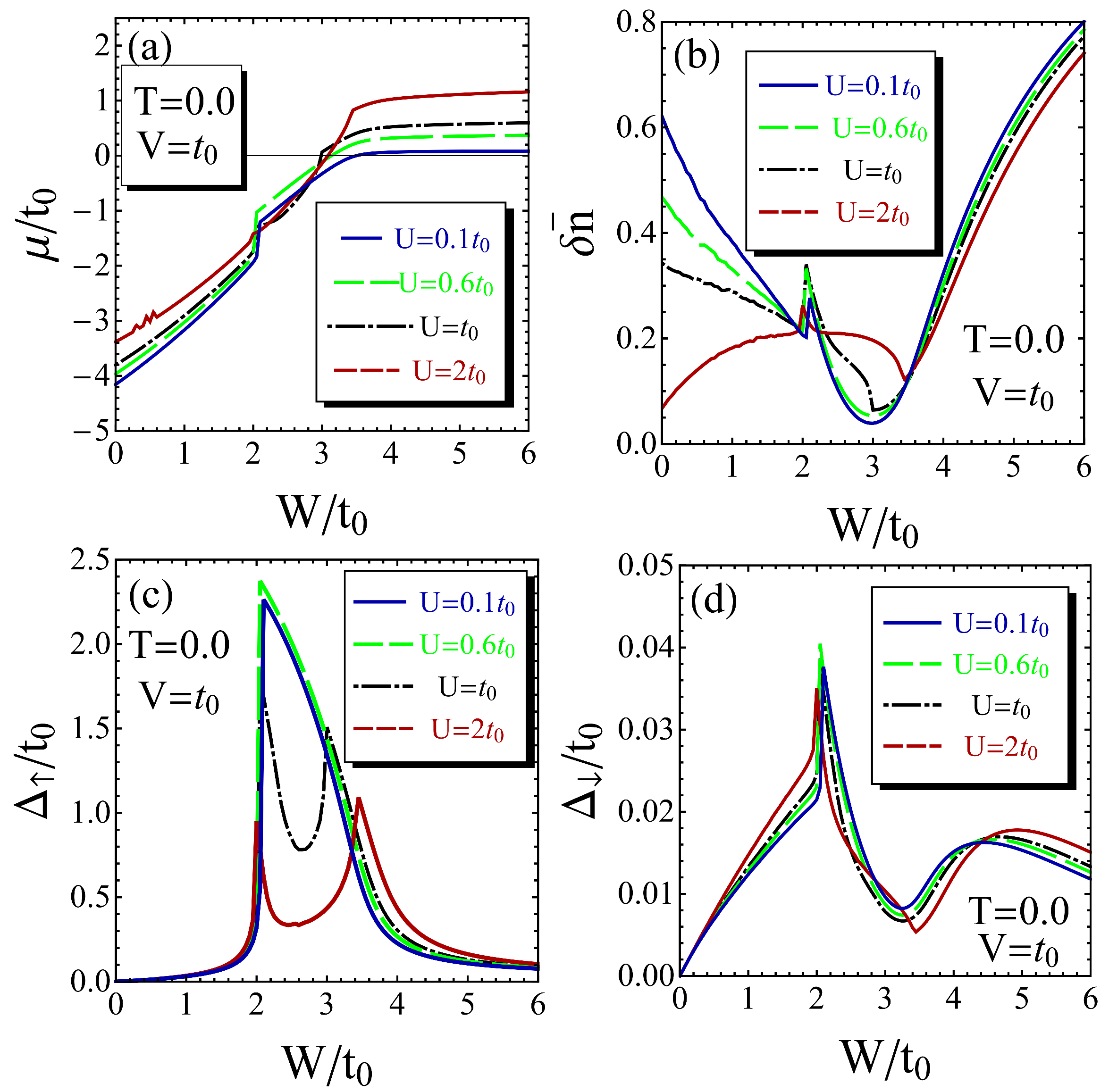}
	\caption{\label{fig:Fig_7}(Color online) The solution of the system of equations in Eq.(\ref{Equation_32}) as a function of the inter-wire Coulomb interaction parameter $W$. The chemical potential (see panel (a)), the average charge density imbalance between wires (see panel (b)), excitonic order parameter $\Delta_{\sigma}$ (see panels (c) and (d)) have been calculated for different values of the local Hubbard-$U$ interaction parameter. The partial-filling was considered with $\kappa=0.8$. The external gate potential and magnetic field parameter have been set at values $V=t_{0}$ and $\tilde{B}=1.0$. The calculations have been done in the zero temperature limit.}
\end{figure}

\begin{figure}
	\includegraphics[scale=0.436]{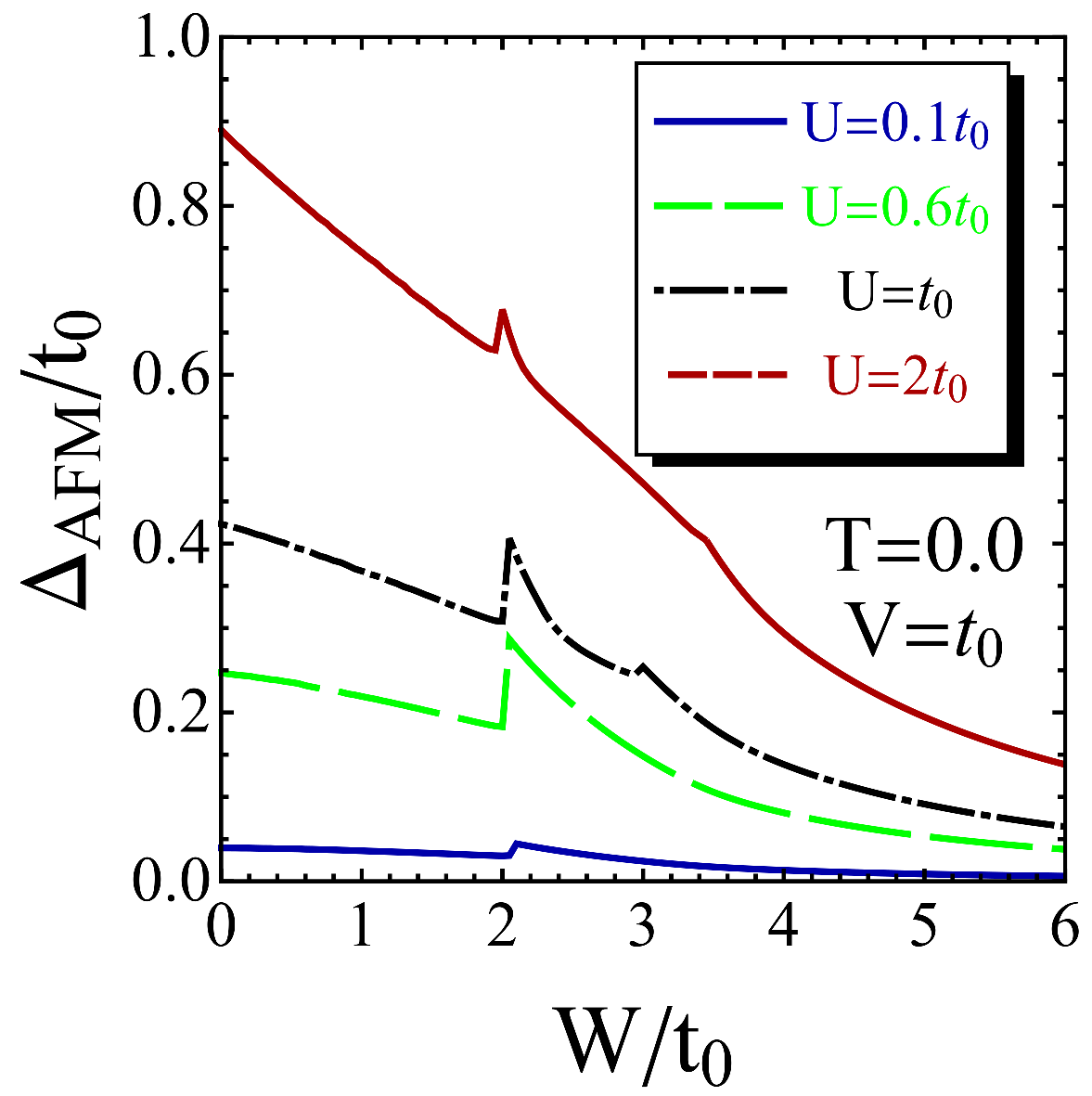}
\caption{\label{fig:Fig_8}(Color online) The antiferromagnetic order parameter as a function of the inter-wire Coulomb interaction parameter $W$, for different values of the Hubbard-$U$ interaction. The plots have been calculated for the fixed field values $V=t_{0}$ and $\tilde{B}=1.0$. The partial-filling case was considered during calculations with $\kappa=0.8$. The system of equations in Eq.(\ref{Equation_32}) has been solved for the zero temperature limit $T=0$.}
\end{figure}

Indeed, the variation of the inter-wire Coulomb interaction parameter $W$ corresponds to the changes of the distance $d$ between wires (see in Fig.~\ref{fig:Fig_1}), and the observation of splitting of the excitonic curves $\Delta_{\uparrow}$, when lowering $\tilde{B}$, could be experimentally observed by Angle Resolved Photoemission Spectroscopy (ARPES) when changing the distance between the layers. At $\tilde{B}<\tilde{B}_{\rm C}$, we have two distinct values of the interaction energy $W_{\rm 01}$ and $W_{\rm 01}$, or two different inter-wire separations $d_{\rm 01}$ and $d_{\rm 02}$ (see left picture in Fig.~\ref{fig:Fig_4}) which gives two strong excitonic pulses for $\sigma=\uparrow$, while, for $\tilde{B}>\tilde{B}_{\rm C}$ we have only one strong excitonic pulse for a given $W_{\rm 0}$ (or $d_{\rm 0}$) (see right picture in Fig.~\ref{fig:Fig_3}). 
%
%
In Fig.~\ref{fig:Fig_4}, we have shown the solutions for the chemical potential (see panel (a)), the average charge density imbalance function $\delta{\bar{n}}$ (see panel (b)) and the antiferromagnetic order parameter in the system $\Delta_{\rm AFM}$ (see panel (c)), from Eq.(\ref{Equation_32}). The considered values of the extermal magnetic field are the same like in Fig.~\ref{fig:Fig_2} above. It is interesting to notice here that at the critical value of the magnetic field $B_{\rm C}$ the anfiterromagnetic order parameter attains its largest value. The other parameters in the system are the same as in Fig.~\ref{fig:Fig_2}. We see in panel (c), in Fig.~\ref{fig:Fig_4}, that for the values $\tilde{B}<\tilde{B}_{\rm C}$ (see plots in red and green in panel (c)) the antiferromagnetic order parameter is decreases at the intermediate values of inter-wire interaction energy $W<2.5t_{0}$. \textit{Indeed, the large antiferromagnetic order in the system for the values $\tilde{B}>\tilde{B}_{\rm C}$ is an artifact of strong localization of the electronic spins. When reducing the magnetic field parameter the electronic spins get delocalized (from the directions they formed along or opposite to the direction of the applied magnetic field) and the corresponding antiferromagnetic order parameter get reduced in this case, which we observe in panel (c), in Fig.~\ref{fig:Fig_4}. }    
%
%
\begin{figure}
	\includegraphics[scale=0.21]{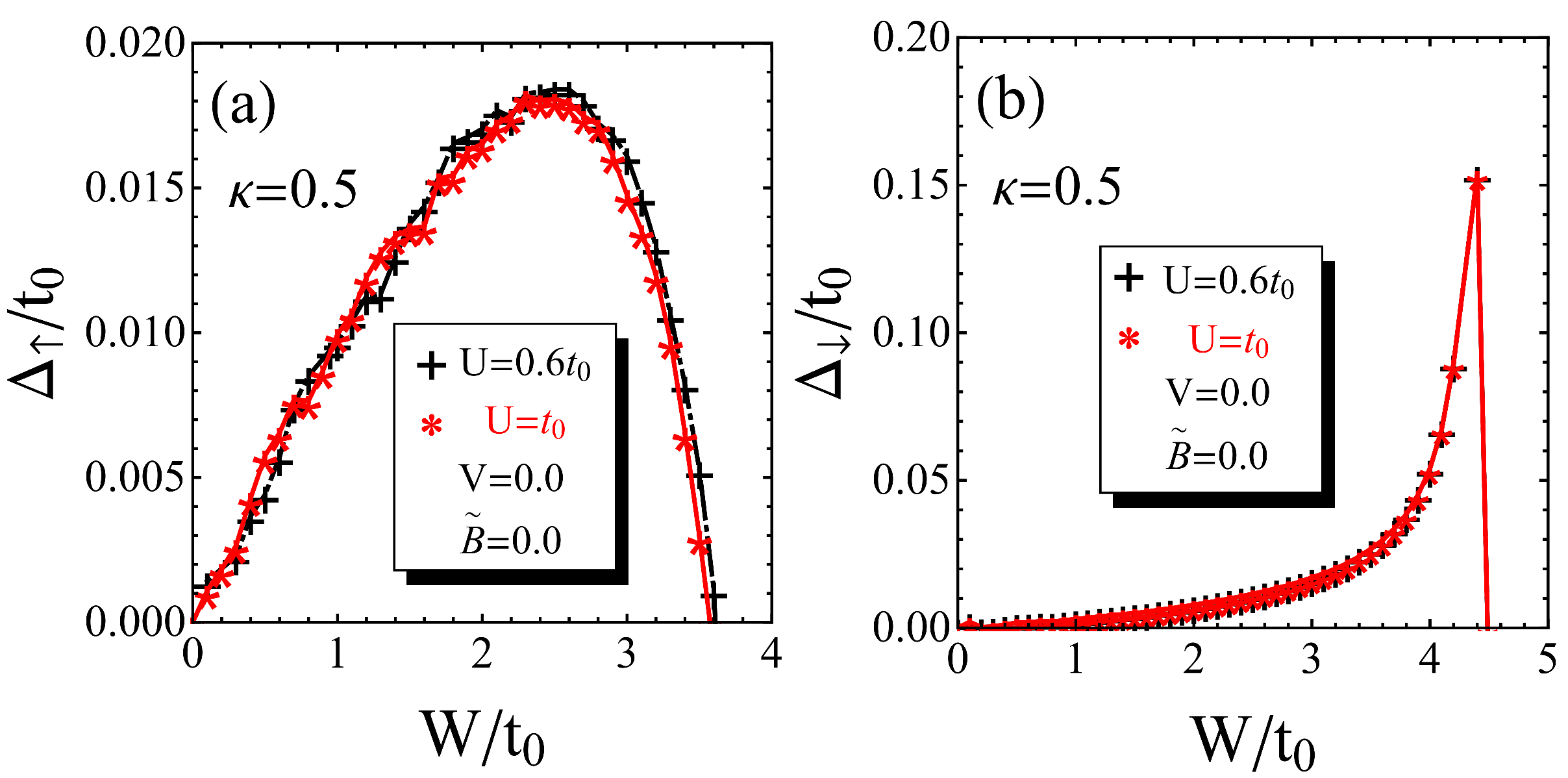}
	\caption{\label{fig:Fig_5}(Color online) The excitonic order parameter $\Delta_{\uparrow}$ (see panel (a)) and $\Delta_{\downarrow}$ (see panel (b)), as a function of the inter-wire Coulomb interaction potential. The case of half-filling has been considered with $\kappa=0.5$. The external gate potential $V$ and magnetic field parameter $\tilde{B}$ have been fixed at values $V=0.0$ and $\tilde{B}=0.0$. Two different values of the Hubbard-$U$ potential have been considered during the calculations. The calculations have been performed at the zero temperature limit $T=0$.}
\end{figure} 
%
The solutions for the case of the half-filling ($\kappa=0.5$) have been shown in Fig.~\ref{fig:Fig_5}. The calculations have been performed for the case of the absence of the external fields $V$ and $B$. Two different values of the Hubbard-$U$ potential have been considered at zero temperature limit. We see that in the case of half-filling the magnitudes of the excitonic order parameters have been gradually reduced and $\Delta_{\downarrow}\gg \Delta_{\uparrow}$, opposite to the case of partial-filling in Fig.~\ref{fig:Fig_2} when $\Delta_{\uparrow}\gg\Delta_{\downarrow}$. Moreover, the behaviors of the excitonic order parameters $\Delta_{\uparrow}$ and $\Delta_{\downarrow}$ are completely different that in the case of partial-filling.   
%
\begin{figure}
	\includegraphics[width=8.5cm, height=9.5cm]{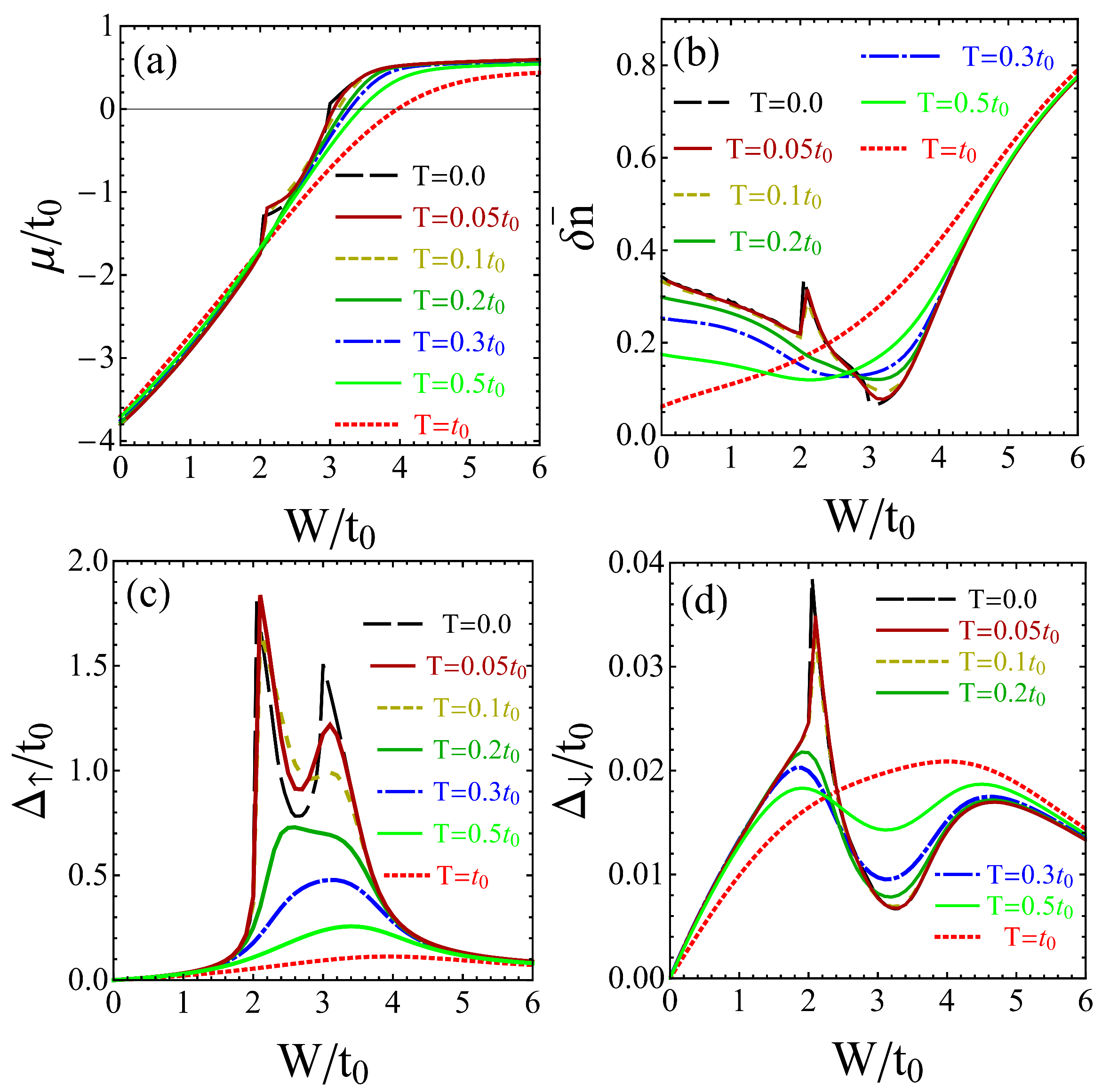}
	\caption{\label{fig:Fig_6}(Color online) The solution of the system of equations in Eq.(\ref{Equation_32}) as a function of the inter-wire Coulomb interaction parameter $W$. The chemical potential (see panel (a)), the average charge density imbalance between wires (see panel (b)), excitonic order parameter $\Delta_{\sigma}$ (see panels (c) and (d)) have been calculated at different temperatures shown in the picture. The partial-filling was considered with $\kappa=0.8$. The external gate potential and magnetic field parameter have been set at values $V=t_{0}$ and $\tilde{B}=1.0$. The Hubbard-$U$ interaction potential has been set at value $U=t_{0}$.}
\end{figure} 
%
The temperature dependence of calculated physical quantities has been shown in Fig.~\ref{fig:Fig_6}. We see in panels (a)-(d) that the significant deviation of curve from their temperature behavior was observed at the temperature $T=0.2t_{0}$. For the inter-site hopping parameter $t_{0}=0.1$ eV, this corresponds to $T=232.1$ K. The chemical potential, at each value of $W$ does not get affected much by the change of temperature, while the other physical quantities such average density imbalance $\delta{\bar{n}}$ (see panel (b)), excitonic order parameters $\Delta_{\uparrow}$, $\Delta_{\downarrow}$ (see panels (c) and (d)) and the antiferromagnetic order parameter $\Delta_{\rm AFM}$ vary significantly with the change of temperature in the system. We observe, particularly, how two peak like structure of the excitonic order parameter $\Delta_{\uparrow}$, shown in panel (c) get smoothed  when increasing temperature up to value $T=0.2t_{0}$ and the magnitude of $\Delta_{\uparrow}$ was gradually decreased at the higher temperatures (see the curves at $T\in\left[0.3t_{0},t_{0}\right]$). The excitonic order parameter with the opposite spin direction shows the opposite behavior. Mainly, one-peak structure get smoothed up to $T=0.2t_{0}$, meanwhile its shape is transforming into two nipple like structure when augmenting the temperature furthermore. Moreover, the magnitude of the function $\Delta_{\downarrow}$ still very large at the high temperature limit (see the plot corresponding to value $T=t_{0}=1160.45$ K, thus near the vicinity of melting points) with assumption that the inter-site hopping $t_{0}$ doesn't changes when increasing temperature.  
In Figs.~\ref{fig:Fig_7} and ~\ref{fig:Fig_8}, we solved the system of equations in Eq.(\ref{Equation_32}), for different limits of the local Hubbard-$U$, responsible for the electron localization at the lattice sites positions. We see, in Fig.~\ref{fig:Fig_7} (see panel (a)) that for the large-$U$ limit, when $U=2t_{0}$, the absolute value of the chemical potential $\left|\right|$ is the largest, practically for all values of the inter-chain Coulomb interaction parameter $W$. Thus, the single-particle excitation quasienergies are largest in this case. This, in turn, prohibit the electron-hole coupled quasiparticles formations and curtails the related energy scales (see the excitonic energy scales $\Delta_{\uparrow}$ for $U=2t_{0}$, in panel (c), in Fig.~\ref{fig:Fig_7}). Surprisingly, the energy scales, related to the excitonic order parameter with opposite direction of spin $\Delta_{\downarrow}$, do not get affected much when augmenting the Hubbard-$U$ potential (see the plots in panel (d), in Figs.~\ref{fig:Fig_7}). Moreover, for the small-$U$ limit (see plots in blue and green, for $U=0.1t_{0}$ and $U=0.6t_{0}$, in panel (c)), there are single single excitonic peaks at some given value of the inter-wire interaction energy $W$, while, for the large values of $U$ (see plots in red and black, in panel (c)), those peaks split into two separated excitonic peaks at some specific values $W_{\rm 01}$ and $W_{\rm 02}$. The similar effect took place in Figs.~\ref{fig:Fig_2}, when varying the magnetic field in the interval $\tilde{B}<\tilde{B}_{\rm C}$. In panel (b), in Fig.~\ref{fig:Fig_7}, we give the numerical results for the average charge density imbalance between wires. The large values of $U$ (see, for example, plots in black and red, corresponding to $U=t_{0}$ and $U=2t_{0}$), localize strongly the electrons on their sites positions and the charge density imbalance function is smaller this case, for the intermediate values of the Coulomb interaction energy $W$. In the large-$W$ limit this behavior remains observable on the plots although $\delta_{\bar{n}}$ is very large in this case, which is the manifestation of the strong charge density fluctuations in the system. Indeed, the large values of $W$ could be achieved, when the wires are two close each other (in other words, when the separation distance is very small), moreover, in this case the charge system get unstable, and leads to strong average electron density fluctuations in both wires. Alongside, for the reason of this, that the excitonic order parameters $\Delta_{\sigma}$ (see in panels (c) and (d)) are gradually decreased in the large-$W$ limit. In addition, those charge density fluctuations are also responsible for the small values of the antiferromagnetic order parameter $\Delta_{\rm AFM}$ in the strong inter-wire interaction limit (see in Fig.~\ref{fig:Fig_8}). As we discussed earlier, the antiferromagnetic order in our system is measure of the electron's localization on their sites positions and the large values of the Hubbard-$U$ parameter favor the AFM order in the system (see Eq.(\ref{Equation_18})). Thus, as we see in Fig.~\ref{fig:Fig_8}, in the small-$W$ limit, the AFM order is strongly enhanced. The energy scales, corresponding to AFM order parameter decrease when diminishing the Hubbard-$U$ interaction parameter.    
%
%
%

\section{\label{sec:Section_5} Conclusions}
%
In this paper, we have considered an extremely interesting problem, related to the formation of singlet excitonic states between two metallic wires, separated from each by a certain distance. Indeed, the pairing between two metallic wires, separated from each other by a certain distance. Indeed, the pairing between the electron and the hole with opposite spin directions (the hole has the opposite spin direction than the electron) is equivalent to propagation of the electron (with a given spin) from one wire to the another. 

We examined the influence of the external electric and magnetic field on the behavior of several, important, physical quantities in the system, like the chemical potential, the average charge density imbalance function between wires and the excitonic order parameter. We have considered different filling regimes for the average number of electrons at the lattice sites positions. We have considered the dependence of calculated physical quantities as a function of the inter-wire Coulomb interaction parameter (normalized to inter-site hopping amplitude in the wires). We have shown that in the weak localization limit, with small value of the Hubbard-$U$ potential, the magnetic field stabilizes the average charge fluctuations in the system and stabilizes the antiferromagnetic ordering in the bi-wire system. At the half-filling regime and at the zero value of the external magnetic field (when the maximum average occupation number of the lattice sites is $1$) we got different behaviors of the excitonic order parameters for different spin directions and the result doesn't changed for different limits of the local Hubbard potential $U$. 

Furthermore, we found temperature dependence of calculated physical quantities at large-$U$ limit and for finite large value of the magnetic field. 

By analyzing those results, we concluded that the magnetic field and temperature could split the one-peak excitonic behaviors, for $\sigma=\uparrow$, into two excitonic pulses at some concrete values $\tilde{B}={\tilde{B}}_{\rm C}$ and $T=T_{\rm C}$. Another, completely different, behavior has been observed for the excitonic order parameter with the opposite spin direction $\sigma=\downarrow$.

We claimed out also that the effect of Hubbard-$U$ interaction on the excitonic order parameters is very similar to the effects of the magnetic field and temperature. Endeavor of the work, we have found that the effect of the Hubbard-$U$ interaction is to change the antiferromagnetic order parameter, and the large values of $U$ increase considerably the antiferromagnetic order parameter, which is the manifestation of strong electron spin-localization on the atomic lattice sites positions, thus, proving the reminiscent artifact that the large-$U$ limit is the strong Mott-Hubbard localization limit.         

The results in the paper could be important especially for the technological applications of the considered system as a system for the information transfer with high velocities and for simultaneous considerations of the coupled excitons as a robust units for transferring the quantum of information in the actually developing quantum information technologies.

\appendix 
%
\section{\label{sec:Section_6} The calculation of important coefficients}
%
In this Section, we present a detailed derivation of the coefficients given, under sums, in Eq.(\ref{Equation_32}). First of all, we write the partition function in Eq.(\ref{Equation_26}) with the source terms
\begin{widetext}
\begin{eqnarray} 
	{\cal{S}}\left[\bar{\psi},\psi\right]=\frac{1}{\beta{N}}\sum_{{\bf{k}}\nu_{n}}\left(-\frac{1}{2}\bar{\psi}({\bf{k}},\nu_{n}){\cal{G}}^{-1}\left({\bf{k}}\nu_{n}\right){\psi}({\bf{k}},\nu_{n})+\frac{1}{2}\bar{J}\left({\bf{k}},\nu_{n}\right)\psi\left({\bf{k}},\nu_{n}\right)+\frac{1}{2}J\left({\bf{k}},\nu_{n}\right)\bar{\psi}\left({\bf{k}},\nu_{n}\right)\right),
	\label{Equation_A_1}
\end{eqnarray}
\end{widetext}
Here, we have introduced the source terms $\bar{J}\left({\bf{k}},\nu_{n}\right)$ and $J\left({\bf{k}},\nu_{n}\right)$ in Nambu forms, analogue to Eqs.(\ref{Equation_24}) and (\ref{Equation_25}).  
Then, a simple Hubbard-Stratonovich transformation gives the following expression for the partition function
\begin{eqnarray}
{\cal{Z}}\approx e^{\frac{1}{2}\sum_{{\bf{k}}},\nu_{n}\bar{J}\left({\bf{k}},\nu_{n}\right){\cal{D}}\left({\bf{k}},\nu_{n}\right)J\left({\bf{k}},\nu_{n}\right)},
\label{Equation_A_2}
\end{eqnarray}
The matrix ${\cal{D}}$ obtained in the right-hand side in the exponential in Eq.(\ref{Equation_A_2}) is the inverse of the matrix in Eqs.(\ref{Equation_{26}}) and (\ref{Equation_27}).
Then, the averages in the problem could be obtained after simple functional differentiation with respect to source terms. For example, we have
\begin{eqnarray}
\frac{\delta^{2}{\cal{Z}}}{\delta{J}_{1\uparrow}\delta{\bar{J}}_{1\uparrow}}=\frac{1}{4}\left\langle\bar{a}_{1\uparrow}a_{1\uparrow}\right\rangle	
\label{Equation_A_3}
\end{eqnarray}
\begin{eqnarray}
\left\langle\bar{a}_{1\uparrow}a_{1\uparrow}\right\rangle=-2{\cal{D}}_{\rm 11}\left({\bf{k}},\nu_{n}\right),
\label{Equation_A_4}
\end{eqnarray}
and we get
\begin{eqnarray}
\left\langle\bar{a}_{1\uparrow}a_{1\uparrow}\right\rangle=-2{\cal{D}}_{\rm 11}\left({\bf{k}},\nu_{n}\right).
\label{Equation_A_5}
\end{eqnarray}
Similarly, we can write
\begin{eqnarray}
\left\langle\bar{a}_{2\uparrow}a_{2\uparrow}\right\rangle=-2{\cal{D}}_{\rm 33}\left({\bf{k}},\nu_{n}\right).	
\label{Equation_A_6}
\end{eqnarray}
The same expression for $\sigma=\downarrow$ have the following form
\begin{eqnarray}
	\left\langle\bar{a}_{1\downarrow}a_{1\downarrow}\right\rangle=-2{\cal{D}}_{\rm 22}\left({\bf{k}},\nu_{n}\right).
	\label{Equation_A_7}
\end{eqnarray}
Concerning the excitonic order parameter, we have the following average
\begin{eqnarray}
\left\langle\bar{a}_{1\uparrow}a_{2\uparrow}\right\rangle=2{\cal{D}}_{\rm 31}\left({\bf{k}},\nu_{n}\right)
\label{Equation_A_8}
\end{eqnarray}
and, for the inverse spin direction, we have
\begin{eqnarray}
	\left\langle\bar{a}_{1\downarrow}a_{2\downarrow}\right\rangle=2{\cal{D}}_{\rm 42}\left({\bf{k}},\nu_{n}\right).  
	\label{Equation_A_9}
\end{eqnarray}
Then, we can write the system of equations for the chemical potential $\mu$, the average charge density imbalance between the wires $\delta{\bar{n}}$, the excitonic order parameters $\Delta_{\uparrow}$ and $\Delta_{\downarrow}$ and the antiferromagnetic order parameter $\Delta_{\rm AFM}$ 
\begin{eqnarray}
&&-\frac{2}{\left(\beta{N}\right)^{2}}\sum_{{\bf{k}},\nu_{n}}\left[{\cal{D}}_{11}\left({\bf{k}},\nu_{n}\right)+{\cal{D}}_{33}\left({\bf{k}},\nu_{n}\right)\right]=\frac{1}{2\kappa},
\nonumber\\
&&-\frac{2}{\left(\beta{N}\right)^{2}}\sum_{{\bf{k}}\nu_{n}}\left[{\cal{D}}_{33}\left({\bf{k}},\nu_{n}\right)-{\cal{D}}_{22}\left({\bf{k}},\nu_{n}\right)\right]=\frac{\delta{\bar{n}}}{2},
\nonumber\\
&&\Delta_{\uparrow}=-\frac{2W}{\left(\beta{N}\right)^{2}}\sum_{{\bf{k}},\nu_{n}}{\cal{D}}_{\rm 31}\left({\bf{k}},\nu_{n}\right),
\nonumber\\
&&\Delta_{\downarrow}=-\frac{2W}{\left(\beta{N}\right)^{2}}\sum_{{\bf{k}}}{\cal{D}}_{\rm 42}\left({\bf{k}},\nu_{n}\right),
\nonumber\\
&&\Delta_{\rm AFM}=-\frac{U}{\left(\beta{N}\right)^{2}}\sum_{{\bf{k}}}\left[{\cal{D}}_{\rm 11}\left({\bf{k}},\nu_{n}\right)-{\cal{D}}_{\rm 22}\left({\bf{k}},\nu_{n}\right)\right].
\nonumber\\
\label{Equation_A_10}
\end{eqnarray}
Furthermore, we can calculate explicitly, the coefficients in the left-hand side in Eq.(\ref{Equation_10}). Namely, for the first equation, in Eq.(\ref{Equation_10}), we obtain
\begin{eqnarray}
{\cal{D}}_{\rm 11}+{\cal{D}}_{\rm 33}=\frac{\beta{N}}{2}\frac{{\cal{P}}^{\left(3\right)}_{\kappa}\left(x\right)}{\det{{\cal{D}}\left(x\right)}},
\label{Equation_A_11}
\end{eqnarray}
where the polynomial ${\cal{P}}^{\left(3\right)}_{\kappa}$ is given in the form 
\begin{eqnarray}
{\cal{P}}^{\left(3\right)}_{\kappa}\left(x\right)=-2x^{3}+a_{1}x^{2}+b_{1}x+c_{1},
\label{Equation_A_12}
\end{eqnarray}
where the parameters $a_{1}$, $b_{1}$ and $c_{1}$ have the following expressions
\begin{eqnarray}
&&a_{1}=-\mu_{1}-2\mu_{2}-\mu_{3}-2\mu_{4},
\nonumber\\
&&b_{1}=2|\Delta_{\downarrow}|^{2}-\mu_{1}\mu_{2}-\mu_{2}\mu_{3}-\mu_{1}\mu_{4}-\mu_{3}\mu_{4}-2\mu_{2}\mu_{4},
\nonumber\\
&&c_{1}=\mu_{1}|\Delta_{\downarrow}|^{2}+\mu_{3}|\Delta_{\downarrow}|^{2}-\mu_{2}\mu_{4}\left(\mu_{1}+\mu_{3}\right).
\label{Equation_A_13}
\end{eqnarray}
For the second equation, in Eq.(\ref{Equation_10}), we obtain
\begin{eqnarray}
{\cal{D}}_{\rm 11}-{\cal{D}}_{\rm 22}=\frac{\beta{N}}{2}\frac{{\cal{P}}^{\left(2\right)}_{\rm AFM}\left(x\right)}{\det{{\cal{D}}\left(x\right)}},
\label{Equation_A_14}
\end{eqnarray}
where the polynomial ${\cal{P}}^{\left(2\right)}\left(x\right)$ is given in the form 
\begin{eqnarray}
{\cal{P}}^{\left(3\right)}\left(x\right)=a_{2}x^{3}+b_{2}x^{2}+c_{2},
\label{Equation_A_15}
\end{eqnarray}
where the parameter $a_{2}$, $b_{2}$ and $c_{2}$ have the following expressions
\begin{eqnarray}
&&a_{2}=\mu_{1}-\mu_{2},
\nonumber\\
&&b_{2}=|\Delta_{\downarrow}|^{2}-|\Delta_{\uparrow}|^{2}+\mu_{3}\left(\mu_{1}-\mu_{2}\right)+\mu_{4}\left(\mu_{1}-\mu_{2}\right),
\nonumber\\
&&c_{2}=|\Delta_{\downarrow}|^{2}\mu_{3}-|\Delta_{\uparrow}|^{2}\mu_{4}+\mu_{3}\mu_{4}\left(\mu_{1}-\mu_{2}\right),
\nonumber\\ 
\label{Equation_A_16}
\end{eqnarray}
\begin{eqnarray}
{\cal{D}}_{\rm 33}-{\cal{D}}_{\rm 22}=\frac{\beta{N}}{2}\frac{{\cal{P}}^{\left(2\right)}_{\rm \delta{\bar{n}}}\left(x\right)}{\det{{\cal{D}}\left(x\right)}},
\label{Equation_A_17}
\end{eqnarray}
and the polynomial ${\cal{P}}^{\left(2\right)}_{\rm \delta{\bar{n}}}$ is the second order polynomial, which is given as
\begin{eqnarray}
{\cal{P}}^{\left(2\right)}_{\rm \delta{\bar{n}}}=a_{3}x^{2}+b_{3}x+c_{3},
\label{Equation_A_18}
\end{eqnarray}
where the parameter $a_{3}$, $b_{3}$ and $c_{3}$ are 
\begin{eqnarray}
&&a_{3}=\mu_{3}-\mu_{2},
\nonumber\\
&&b_{3}=|\Delta_{\downarrow}|^{2}-|\Delta_{\uparrow}|^{2}-\mu_{1}\left(\mu_{2}-\mu_{3}\right),
\nonumber\\
&&c_{3}=\mu_{1}|\Delta_{\downarrow}|^{2}-\mu_{4}|\Delta_{\uparrow}|^{2}-\mu_{1}\mu_{4}\left(\mu_{2}-\mu_{3}\right).
\label{Equation_A_19}
\end{eqnarray}
Next, the coefficients ${\cal{D}}_{\rm 31}$ and ${\cal{D}}_{\rm 42}$ in the equations for the excitonic order parameters $\Delta_{\uparrow}$ and $\Delta_{\downarrow}$ in Eq.(\ref{Equation_A_10}) are
\begin{eqnarray}
{\cal{D}}_{\rm 31}=\frac{\beta{N}}{2}\frac{{\cal{P}}^{\left(2\right)}_{\Delta_{\uparrow}}\left(x\right)\left(\Delta_{\uparrow}+\gamma_1\right)}{\det{{\cal{D}}\left(x\right)}},
\label{Equation_A_20}
\end{eqnarray}
\begin{eqnarray}
{\cal{D}}_{\rm 42}=\frac{\beta{N}}{2}\frac{{\cal{P}}^{\left(2\right)}_{\Delta_{\downarrow}}\left(x\right)\left(\Delta_{\downarrow}+\gamma_1\right)}{\det{{\cal{D}}\left(x\right)}},
\label{Equation_A_21}
\end{eqnarray}
where the polynomials ${\cal{P}}^{\left(2\right)}_{\rm \Delta_{\uparrow}}\left(x\right)$ and ${\cal{P}}^{\left(2\right)}_{\rm \Delta_{\downarrow}}\left(x\right)$ are the second order polynomials
\begin{eqnarray}
{\cal{P}}^{\left(2\right)}_{\rm \Delta_{\uparrow}}=x^{2}+b_{4}x+c_{4},
\nonumber\\
{\cal{P}}^{\left(2\right)}_{\rm \Delta_{\downarrow}}=x^{2}+b_{5}x+c_{5},
\label{Equation_A_22}
\end{eqnarray}
where the coefficients $b_{4}$, $c_{4}$, $b_{5}$ and $c_{5}$ have been introduced as
\begin{eqnarray}
&&b_{4}=\mu_{2}+\mu_{4},
\nonumber\\
&&c_{4}=\mu_{2}\mu_{4}-|\Delta_{\downarrow}|^{2},
\nonumber\\
&&b_{5}=\mu_{1}+\mu_{3},
\nonumber\\
&&c_{5}=\mu_{1}\mu_{3}-|\Delta_{\uparrow}|^{2}.
\label{Equation_A_23}
\end{eqnarray}
Remeber that the equation for the determinant
\begin{eqnarray}
\det{\cal{D}}\left(x\right)=\left(x-\varepsilon_{1}\right)\left(x-\varepsilon_{2}\right)\left(x-\varepsilon_{3}\right)\left(x-\varepsilon_{4}\right)=0
\nonumber\\
\label{Equation_A_24}
\end{eqnarray}
gives the energy spectrum of the problem.
Aftermore, by evaluating the fractions in Eqs.(\ref{Equation_A_11}), (\ref{Equation_A_14}), (\ref{Equation_A_17}), (\ref{Equation_A_20}) and (\ref{Equation_A_21}), we can write the system of self-consistent equations in Eq.(\ref{Equation_A_10}) in the following compact form
\begin{eqnarray}
&&-\frac{1}{\beta{N}}\sum^{4}_{i=1}\sum_{{\bf{k}},\nu_{n}}\frac{\alpha_{i{\bf{k}}}}{-i\nu_{n}-\varepsilon_{i{\bf{k}}}}=\frac{1}{2\kappa},
\nonumber\\
&&-\frac{1}{\beta{N}}\sum^{4}_{i=1}\sum_{{\bf{k}},\nu_{n}}\frac{\beta_{i{\bf{k}}}}{-i\nu_{n}-\varepsilon_{i{\bf{k}}}}=\frac{\delta{\bar{n}}}{2},
\nonumber\\
&&\Delta_{\uparrow}=-W\frac{\Delta_{\uparrow}+\gamma_1}{\beta{N}}\sum^{4}_{i=1}\sum_{{\bf{k}},\nu_{n}}\frac{\delta^{\left(1\right)}_{i{\bf{k}}}}{-i\nu_{n}-\varepsilon_{i{\bf{k}}}},
\nonumber\\
&&\Delta_{\downarrow}=-W\frac{\Delta_{\uparrow}+\gamma_1}{\beta{N}}\sum^{4}_{i=1}\sum_{{\bf{k}},\nu_{n}}\frac{\delta^{\left(2\right)}_{i{\bf{k}}}}{-i\nu_{n}-\varepsilon_{i{\bf{k}}}},
\nonumber\\
&&\Delta_{\rm AFM}=-\frac{U}{2\beta{N}}\sum^{4}_{i=1}\sum_{{\bf{k}},\nu_{n}}\frac{\gamma_{i{\bf{k}}}}{-i\nu_{n}-\varepsilon_{i{\bf{k}}}},
\label{Equation_A_25}
\end{eqnarray}
where the coefficients $\alpha_{i{\bf{k}}}$, $\beta_{i{\bf{k}}}$, $\gamma_{i{\bf{k}}}$, $\delta^{\left(1\right)}$ and $\delta^{\left(2\right)}$ are given in the following forms
\begin{widetext}
	\begin{eqnarray}
		\footnotesize
		\arraycolsep=0pt
		\medmuskip = 0mu
		\alpha_{i{\bf{k}}}
		=(-1)^{i+1}
		\left\{
		\begin{array}{cc}
			\displaystyle  & \frac{{\cal{P}}^{\left(3\right)}_{\kappa}\left(\epsilon_{i\sigma}\left({\bf{k}}\right)\right)}{\epsilon_{1\sigma}\left({\bf{k}}\right)-\epsilon_{2\sigma}\left({\bf{k}}\right)}\prod_{j=3,4}\frac{1}{\epsilon_{i\sigma}\left({\bf{k}}\right)-\epsilon_{j\sigma}\left({\bf{k}}\right)},  \ \ \  $if$ \ \ \ i=1,2,
			\newline\\
			\newline\\
			\displaystyle  & \frac{{\cal{P}}^{\left(3\right)}_{\kappa}\left(\epsilon_{i\sigma}\left({\bf{k}}\right)\right)}{\epsilon_{3\sigma}\left({\bf{k}}\right)-\epsilon_{4\sigma}\left({\bf{k}}\right)}\prod^{}_{j=1,2}\frac{1}{\epsilon_{i\sigma}\left({\bf{k}}\right)-\epsilon_{j\sigma}\left({\bf{k}}\right)},  \ \ \  $if$ \ \ \ i=3,4,
		\end{array}\right.
		\nonumber\\
		\label{Equation_A_26}
	\end{eqnarray}
	\begin{eqnarray}
		\footnotesize
		\arraycolsep=0pt
		\medmuskip = 0mu
		\beta_{i{\bf{k}}}
		=(-1)^{i+1}
		\left\{
		\begin{array}{cc}
			\displaystyle  & \frac{{\cal{P}}^{\left(2\right)}_{\kappa}\left(\epsilon_{i\sigma}\left({\bf{k}}\right)\right)}{\epsilon_{1\sigma}\left({\bf{k}}\right)-\epsilon_{2\sigma}\left({\bf{k}}\right)}\prod_{j=3,4}\frac{1}{\epsilon_{i\sigma}\left({\bf{k}}\right)-\epsilon_{j\sigma}\left({\bf{k}}\right)},  \ \ \  $if$ \ \ \ i=1,2,
			\newline\\
			\newline\\
			\displaystyle  & \frac{{\cal{P}}^{\left(2\right)}_{\kappa}\left(\epsilon_{i\sigma}\left({\bf{k}}\right)\right)}{\epsilon_{3\sigma}\left({\bf{k}}\right)-\epsilon_{4\sigma}\left({\bf{k}}\right)}\prod^{}_{j=1,2}\frac{1}{\epsilon_{i\sigma}\left({\bf{k}}\right)-\epsilon_{j\sigma}\left({\bf{k}}\right)},  \ \ \  $if$ \ \ \ i=3,4,
		\end{array}\right.
		\nonumber\\
		\label{Equation_A_27}
	\end{eqnarray}
		\begin{eqnarray}
		\footnotesize
		\arraycolsep=0pt
		\medmuskip = 0mu
		\delta^{\left(1\right)}_{i{\bf{k}}}
		=(-1)^{i+1}
		\left\{
		\begin{array}{cc}
			\displaystyle  & \frac{{\cal{P}}^{\left(2\right)}_{\Delta_{\uparrow}}\left(\epsilon_{i\sigma}\left({\bf{k}}\right)\right)}{\epsilon_{1\sigma}\left({\bf{k}}\right)-\epsilon_{2\sigma}\left({\bf{k}}\right)}\prod_{j=3,4}\frac{1}{\epsilon_{i\sigma}\left({\bf{k}}\right)-\epsilon_{j\sigma}\left({\bf{k}}\right)},  \ \ \  $if$ \ \ \ i=1,2,
			\newline\\
			\newline\\
			\displaystyle  & \frac{{\cal{P}}^{\left(2\right)}_{\Delta_{\uparrow}}\left(\epsilon_{i\sigma}\left({\bf{k}}\right)\right)}{\epsilon_{3\sigma}\left({\bf{k}}\right)-\epsilon_{4\sigma}\left({\bf{k}}\right)}\prod^{}_{j=1,2}\frac{1}{\epsilon_{i\sigma}\left({\bf{k}}\right)-\epsilon_{j\sigma}\left({\bf{k}}\right)},  \ \ \  $if$ \ \ \ i=3,4,
		\end{array}\right.
		\nonumber\\
		\label{Equation_A_28}
	\end{eqnarray}
	\begin{eqnarray}
		\footnotesize
		\arraycolsep=0pt
		\medmuskip = 0mu
		\delta^{\left(2\right)}_{i{\bf{k}}}
		=(-1)^{i+1}
		\left\{
		\begin{array}{cc}
			\displaystyle  & \frac{{\cal{P}}^{\left(2\right)}_{\Delta_{\downarrow}}\left(\epsilon_{i\sigma}\left({\bf{k}}\right)\right)}{\epsilon_{1\sigma}\left({\bf{k}}\right)-\epsilon_{2\sigma}\left({\bf{k}}\right)}\prod_{j=3,4}\frac{1}{\epsilon_{i\sigma}\left({\bf{k}}\right)-\epsilon_{j\sigma}\left({\bf{k}}\right)},  \ \ \  $if$ \ \ \ i=1,2,
			\newline\\
			\newline\\
			\displaystyle  & \frac{{\cal{P}}^{\left(2\right)}_{\Delta_{\downarrow}}\left(\epsilon_{i\sigma}\left({\bf{k}}\right)\right)}{\epsilon_{3\sigma}\left({\bf{k}}\right)-\epsilon_{4\sigma}\left({\bf{k}}\right)}\prod^{}_{j=1,2}\frac{1}{\epsilon_{i\sigma}\left({\bf{k}}\right)-\epsilon_{j\sigma}\left({\bf{k}}\right)},  \ \ \  $if$ \ \ \ i=3,4,
		\end{array}\right.
		\nonumber\\
		\label{Equation_A_29}
	\end{eqnarray}
	\begin{eqnarray}
		\footnotesize
		\arraycolsep=0pt
		\medmuskip = 0mu
		\gamma_{i{\bf{k}}}
		=(-1)^{i+1}
		\left\{
		\begin{array}{cc}
			\displaystyle  & \frac{{\cal{P}}^{\left(2\right)}_{\Delta_{\rm AFM}}\left(\epsilon_{i\sigma}\left({\bf{k}}\right)\right)}{\epsilon_{1\sigma}\left({\bf{k}}\right)-\epsilon_{2\sigma}\left({\bf{k}}\right)}\prod_{j=3,4}\frac{1}{\epsilon_{i\sigma}\left({\bf{k}}\right)-\epsilon_{j\sigma}\left({\bf{k}}\right)},  \ \ \  $if$ \ \ \ i=1,2,
			\newline\\
			\newline\\
			\displaystyle  & \frac{{\cal{P}}^{\left(2\right)}_{\Delta_{\rm AFM}}\left(\epsilon_{i\sigma}\left({\bf{k}}\right)\right)}{\epsilon_{3\sigma}\left({\bf{k}}\right)-\epsilon_{4\sigma}\left({\bf{k}}\right)}\prod^{}_{j=1,2}\frac{1}{\epsilon_{i\sigma}\left({\bf{k}}\right)-\epsilon_{j\sigma}\left({\bf{k}}\right)},  \ \ \  $if$ \ \ \ i=3,4,
		\end{array}\right.
		\nonumber\\
		\label{Equation_A_30}
	\end{eqnarray}
	\end{widetext}
Then, we perform the fermionic Matsubara summation over the frequencies $\nu_{n}$ and we rewrite the system of self-consistent equationsin the following form
\begin{eqnarray}
&&-\frac{1}{N}\sum_{{\bf{k}}}\sum^{4}_{i=1}\alpha_{i{\bf{k}}}n_{\rm F}\left[\varepsilon_{i{\bf{k}}}\right] =\frac{1}{2\kappa},
\nonumber\\
&&-\frac{1}{N}\sum_{{\bf{k}}}\sum^{4}_{i=1}\beta_{i{\bf{k}}}n_{\rm F}\left[\varepsilon_{i{\bf{k}}}\right]=\frac{\delta{\bar{n}}}{2},
\nonumber\\
&&\Delta_{\uparrow}=-\frac{2W}{N}\sum^{4}_{i=1}\delta^{\left(1\right)}_{i{\bf{k}}}n_{\rm F}\left[\varepsilon_{i{\bf{k}}}\right],
\nonumber\\
&&\Delta_{\downarrow}=-\frac{2W}{N}\sum^{4}_{i=1}\delta^{\left(2\right)}_{i{\bf{k}}}n_{\rm F}\left[\varepsilon_{i{\bf{k}}}\right],
\nonumber\\
&&\Delta_{\rm AFM}=-\frac{U}{2N}\sum_{{\bf{k}}}\gamma_{i{\bf{k}}}n_{\rm F}\left[\varepsilon_{i{\bf{k}}}\right].
\nonumber\\
\label{Equation_A_10}
\end{eqnarray}
The explicit expression of the energy spectrum of two metallic wire system in interaction is discussed above, in the Section \ref{sec:Section_3_2}, in Eq.(\ref{Equation_29}).
 
\section*{References}
%
\bibliography{references_authors}

\end{document}